\documentclass[eqsecnum,aps,prd,nofootinbib,twocolumn]{revtex4}

\usepackage{enumerate}
\begin{document}

\def\beq{\begin{equation}}
\def\endeq{\end{equation}}
\def\bea{\begin{eqnarray}}
\def\endea{\end{eqnarray}}

\def\journaldata#1#2#3#4{{\it #1 } {\bf #2:} #3 (#4)}
\def\eprint#1{$\langle$#1\hbox{$\rangle$}}
\def\lto{\mathop
        {\hbox{${\lower3.8pt\hbox{$<$}}\atop{\raise0.2pt\hbox{$\sim$}}$}}}

\title{Resource Letter NSST-1: The Nature and Status of String Theory}

\author{Donald Marolf}

\affiliation{Department of Physics, UCSB, Santa Barbara, CA 93106.}

\begin{abstract}
This Resource Letter provides a guide to some of the the introductory and review 
literature in string theory. It is in no way complete, though it is intended to be of use to students at several levels.  Owing to the nature of the subject, even
much of the introductory literature is quite technical by the
standards of many Resource Letters, requiring prior knowledge of
quantum field theory and general relativity. This Resource
Letter is thus somewhat different from others.  The first part 
describes a few more popular accounts of string theory, which are primarily addressed to the general public, but those with an understanding of basic physics 
will be able to read them more deeply and so obtain a useful rough
orientation to the field. The second part describes resources that are available at the advanced undergraduate level, and the balance describes string
resources for more advanced students.  The latter ranges from general introductions to recent review articles
on branes and black holes, gauge/gravity duality, string field theory, non-commutative geometry, non-BPS branes, tachyon condensation, phenomenology, brane worlds, orbifolds, Calabi-Yau manifolds, and holography.
\end{abstract}
\maketitle

\section{Introduction}

String theory contains an exciting set of ideas that could realize
long-standing hopes for the unification of forces, an
understanding of quantum gravity, and other fundamental issues.
The field is vast and often highly technical.  I attempt to provide below 
a rough guide to the introductory and
review literature, from whence the advanced student can proceed to
forefront research papers.

My goal in this letter is {\it not} to provide a step-by-step
guide leading the student from the most basic level through to
modern topics.  Such a program would require several years of study
by which time the advanced literature will have changed
significantly. Rather, my intention is to provide references be of use to
students at each of several different levels who wish to learn more about string theory. 
References labeled
(E) elementary are accessible to those without training in quantum
mechanics and, typically, to those without any training in physics
at all. These are mostly at the so-called ``popular"
level, intended to convey a few interesting ideas and a very broad
overview to a wide audience. In practice, some high school or
undergraduate training in physics is useful for following them in detail. I give these references in section
\ref{ele}.

Recently a useful set of resources has become available at the advanced undergraduate level (in the sense of the  educational system in the U.S.), and in particular that require no prior knowledge of quantum field theory.  I have rated these (I) intermediate.  While they can not bring the reader up to the level of current research, they provide a useful overview of many topics in more depth than is possible at the purely popular level.  Owing to the vast difference in level between these and either the more advanced or less advanced resources, I have collected most references at this level in section \ref{ug}.  Readers interested in this level of material are well advised to focus on these resources.

Most technical discussions of string theory assume
prior knowledge of quantum field theory (sometimes at an advanced
level) and often assume general relativity as well.  I have rated such works (A)
advanced below.  These advanced resources further break into two categories.
Introductions and general overviews of string theory (mainly
textbooks) are listed in section \ref{gen}, which
also contains a general description of the broader set of
available resources.   More advanced special topics assume
exposure to or knowledge of such introductory material, and I have
listed these in the following sections by topic. Topics included here
are branes and black holes (section \ref{first}), gauge/gravity
duality (section \ref{dual}), string field theory, non-commutative
geometry, non-BPS branes, and tachyon condensation (section
\ref{nonBPS}),  phenomenology, brane worlds, and orbifolds (section \ref{braneworld}), Mirror Symmetry and Calabi-Yau manifolds (section \ref{math}), and the holographic
principle (section \ref{last}).

In each category below, I have attempted to keep the list of references short to avoid large-scale duplication of material. 
I therefore have emphasized broad reviews instead of the original literature, and the reader should be aware that their authors did not necessarily originate all of the work they describe.  The interested reader can of course find the original literature through these reviews.   In each case, I have attempted to include the resources most centrally related to the topic and on which I have received the most positive reports from users.

\section{Popular Material}
\label{ele}

Since the string-theory literature quickly becomes highly technical, I have separated the material that does not require a knowledge of advanced physics and placed it in this section.  All items here are ``elementary" by the criteria described in the introduction, and are useful references for all levels.
Undergraduate students also tend to enjoy this material and can learn much from it.

\subsection{Popular Books}

The two books below seem to be the favorite popular accounts of string theory among undergraduates students.

\begin{enumerate}[{\bf 1{.}}]

\item {\bf The Elegant Universe: Superstrings, Hidden Dimensions, and the Quest for the Ultimate Theory}, B. Greene (W.W. Norton, New York, 1999). Certainly the most popular book on the subject.  The first part gives a basic overview of strings as well as background material on relativity and quantum mechanics.  The second part describes more advanced ideas concerning details of the extra dimensions required by string theory.  (E)

\item {\bf Hyperspace: A Scientific Odyssey Through Parallel Universes, Time Warps and the Tenth Dimension}, M. Kaku (Oxford University Press, New York, 1994). This work focuses on the extra dimensions of string theory, again including much background material. (E)

\end{enumerate}

\subsection{Web Sites}

I highly recommend the web sites below.

\begin{enumerate}[{\bf 1{.}}]

\addtocounter{enumi}{2}

 \item \label{patriciaE} ``The Official String Theory Web
Site -- basic version," P. Schwarz, http://superstringtheory.com/. This
excellent site addresses string theory at many levels and is
appropriate for a broad audience.  The basic version presents a broad overview 
without equations, though a more advanced treatment (see (\ref{patriciaI}))
is available at the same URL.  The reader can also learn about
the history of the subject and some of the people involved.  Physics
topics addressed include Basics of Strings, Experiments,
Cosmology, and Black Holes. (E)

\item ``Beyond String Theory," J. Troost, http://tena4.vub.ac.be/beyondstringtheory/.
This site is still under construction, but the bits currently in place are promising.  The level is somewhat between the basic and advanced levels of ``The Official String Theory Web Site," but again the target audience is quite broad.  In addition to topics such as accelerators, relativity, string theory, unification, banes, M-theory, and so forth, the site also includes a number of ``Intermezzos" on sociology, nobel prizes, and the like. (E)

\item ``Superstrings! Home Page," J. Pierre, http://www.sukidog.com/jpierre/strings/.
The on-line tutorial here has been very popular for some time.  (E)

\item ``A World of Strings," B. Levy,  E. Servam Schreiber, and B. Duhay, http://www.hyper-mind.com/hypermind/universe/content/gsst.htm.  This site describes the very basics of string theory through excerpts from interviews with John Schwarz and David Gross, two of the leaders in the field.  (E)

\item ``NOVA: The elegant Universe,'' Public Broadcasting Service,  http://www.pbs.org/wgbh/nova/elegant/.  This site includes a variety of resources related to the NOVA program ``The elegant Universe'' which aired on PBS Oct. 28 and Nov. 4.  All three hours of the NOVA broadcast can be watched online.

\end{enumerate}

\subsection{Popular Articles}

The reader can find a number of interesting string theory articles in popular-science publications
such as  {\it Scientific American}, {\it Physics World}, {\it Physics Today}, {\it New Scientist} and, at a slightly more advanced level,  {\it Science} and {\it Nature}.  Some fairly recent articles are listed below, but more no doubt will appear in the near future.

\begin{enumerate}[{\bf 1{.}}]

\addtocounter{enumi}{7}

\item ``Explaining everything,"
M. Mukerjee,  Scientific American {\bf 274} (1), 88-94 (January 1996).  This is a good introduction to the basics of string theory, as well as to the more modern topic of ``duality" that has shown surprising links between what were once viewed as different theories of strings.    Some relations between string physics and black holes are explained. (E)

\item ``Particle Physics: the next generation," J. Ellis, Physics World {\bf 12} (12), 43-48 (December 1999).
This article describes particle physicists' hopes for new physics, in the regime known as ``beyond the standard model."   The last part of the article explains how string theory fits into these hopes.  (E)

\item ``Desperately Seeking Superstrings,'' P. Ginsparg and S. Glashow, 
Physics Today {\bf 39} (5), 7-9 (May 1986) [arXiv:physics/9403001].  Ginsparg and Glashow provide
a popular-level critique of string theory as of 1986.  Though now somewhat
dated, many of the comments are still relevant to string theory today, providing a sober
contrast to most other popular accounts.
(E)

\item  ``String Theory: a theory in search of an experiment,'' H. J. Schnitzer, arxiv:physics/0311047.  Another sober discussion of string theory which comments on many modern developments.  While written at the elementary level, many references are made to more advanced concepts. (E)

\item ``Unity from Duality," P. Townsend,  Physics World {\bf 8} (9),  (September 1995).   Here one of the experts explains how the concept of ``duality" has brought unity to what once seemed like a somewhat fragmented world of strings. (E)

\item ``Reflections on the Fate of Spacetime," E. Witten , Physics Today {\bf 49} (4), 24-30 (April 1996). This commentary by a leader in the field describes his thoughts on string theory's implications for the nature of space and time.  (E)

\item ``String Theorists Find a Rosetta Stone,'' G. Taubes, Science {\bf 285}, 512-517
(5247) (1999).  Taubes provides a nice perspective on both successes and difficulties
of string theory and explains the so-called "AdS/CFT duality." (E)

\item ``Into the eleventh dimension," M. Kaku, New Scientist
{\bf 153} (2065), 32-36  (1997). The author of {\it Hyperspace}
writes about M-theory, which introduces an eleventh dimension in
addition to the ten already present in string theory.  (E)

\item \label{hollow}
``The hollow universe," J. R. Minkel, New Scientist {\bf 174},
(2340),  22-27, (2002).  Minkel provides an in-depth but popular-level look at a set of ideas known as ``the holographic principle,"
which some researchers expect to play a fundamental role in string
theory. (E)

\item  ``Strung Out on the Universe: Interview with Raphael
Bousso," J. R. Minkel, Scientific American.com, 3 pages, (April 7, 2003).  This
interesting interview with a leading young researcher discusses
string theory, dark energy, and the holographic principle.  Available only in electronic form.  Search http://www.sciam.com/ by title.  (E)

\item  ``Information in the holographic Universe," J. Bekenstein, Scientific American
{\bf 289} (8), 58-65.  Another popular discussion of holographic ideas.
Bekenstein is an advocate of the ideas discussed, and this article is correspondingly less
skeptical than ref. (\ref{hollow}). (E)

\item ``Brane New Worlds,'' J. P. Gauntlett, Nature {\bf 404},  28-29 (2000).
Gauntlett provides an accessible summary of one version of the idea that our universe may be only a thin "membrane" in some higher
dimensional space and describes why this idea is of interest in particle physics. (E)
\end{enumerate}

\section{Advanced Undergraduate Material}

In this section I describe resources at the advanced (U.S.) undergraduate level; i.e., which requires no prior exposure to quantum field theory.  I begin with a useful web site and then proceed to a list of review articles giving overviews of string theory as a whole (ref. \ref{jerome}) or of various modern topics (refs. \ref{JS}-\ref{SH}).   Most of these articles came from a special section in volume 89 of {\it Current Science}, edited by Spenta Wadia as an outgrowth of the Strings 2001 conference at the Tata Institute of Fundamental Research, Mumbai, India, from January 5-10, 2001.  Most of these works summarize special topics that are treated in
more detail in sections \ref{first} - \ref{last}.  However, I have collected them here for the benefit of students at this level. Together with (ref. \ref{jerome}), they provide a useful overview of string theory as of 2001.  In part \ref{ugbook}, I then describe the one undergraduate-level string theory textbook.

\label{ug}

\subsection{Web Sites}
\label{ugweb}

\begin{enumerate}[{\bf 1{.}}]

\addtocounter{enumi}{19}

 \item \label{patriciaI} ``The Official String Theory Web
Site -- advanced version," P. Schwarz, http://superstringtheory.com/. The ``advanced'' version of 
(ref. \ref{patriciaE}) presents an overview of string theory and related topics at the U.S. advanced undergraduate level.
Strings, black holes, particle theory and experiment, and cosmology are discussed.
The reader can also learn about
the history of the subject and some of the people involved. (I)
\end{enumerate}

\subsection{Articles}
\label{ugart}

\begin{enumerate}[{\bf 1{.}}]

\addtocounter{enumi}{20}

\item \label{jerome} ``M Theory: Strings, Duality, and Branes,'' J.P. Gauntlett, Contemp. Phys. {\bf 39}, 317-328 (1998).
Gauntlett provides an excellent overview of string theory at the advanced U.S. undergraduate level.  The reader should have
a working knowledge of power series, wave equations and separation of variables, electromagnetism, and basic quantum mechanics.
Students are then introduced
to many useful concepts such as perturbation theory, Grand Unified theories, supersymmetry, Kaluza-Klein theory, dualities, and non-perturbative effects.  A postscript version of this article is currently available at 
http://www.strings.ph.qmw.ac.uk/WhatIs/jerome.ps.  (I)

\item \label{JS} ``String Theory,'' J. H. Schwarz, Current Science {\bf 81} (12), 1547-1553 (2001).  Schwarz reviews the history of string theory and introduces the basic idea of string perturbation theory.
(I)

\item \label{TY} ``String Theory and the uncertainty principle,'' T. Yoneya, Current Science {\bf 81} (12), 1554-1560 (2001).  Yoneya describes thoughts on how string theory may modify the uncertainty principle.
(I)

\item \label{AS} ``String Theory and Tachyons,'' A. Sen, Current Science {\bf 81} (12), 1561-1567 (2001).  After a brief review of string theory and dualities, Sen describes the investigation of tachyons in string theory.  Note that the term "tachyon" as used here does not imply faster than light communication, but merely denotes an instability of some state or set of states.  (I)

\item \label{RG} ``Geometry and String Theory,'' R. Gopakumar, Current Science {\bf 81} (12), 1568-1575 (2001).  Strings and branes interact with spacetime geometry in a more subtle way than do point particles.  Gopakumar describes this interaction and its relevance to various dualities, including issues associated with changing the topology (i.e., the connectivity) of spacetime.
(I)

\item \label{EW} ``Black Holes and Quark Confinement,'' E. Witten, Current Science {\bf 81} (12), 1576-1581 (2001).  Witten takes the reader on a tour of a number of ideas, ranging from connections of strings to quarks to black holes and the AdS/CFT duality.  The climax is the fusion of these ideas into a stringy derivation of quark confinement in theories mathematically similar to, but still different than, the QCD that describes the strong interactions of our universe. (I)

\item \label{ST} ``Holography, Black Holes, and String Theory,'' S. P. Trivedi, Current Science {\bf 81} (12), 1582-1590 (2001).
Trivedi discusses the idea that physics might be described by a theory in {\it less} dimensions than we appear to observe.
Known as "holography" due to its analogy with the capture of three-dimensional images on two-dimensional film, this idea is discussed in two forms:  first in its general (and more controversial) form as connected with general issues of black hole entropy, and then in its specific implementation in terms of the so-called "AdS/CFT duality."
(I)

\item \label{SW} ``A Microscopic Theory of Black Holes in String Theory,'' S. R. Wadia, Current Science {\bf 81} (12), 1591-1597 (2001).
After providing an undergraduate-level description of black-hole thermodynamics and Hawking radiation, Wadia describes
the extent to which these phenomena are currently described by string theory.  The concluding section includes a useful set of open questions.  (I)

\item \label{AD} ``Non-critical String Theory,'' A. Dhar, Current Science {\bf 81} (12), 1598-1608 (2001).  Dhar reviews the basics of string theory and discusses string theory as a two-dimensional theory of quantum gravity.  Though it is usually stated that string theory requires ten dimensions, this is the case only if the dimensions are to have certain familiar properties.  While so-called "non-critical strings" outside of ten dimensions will not be able to describe the universe we see, they provide an interesting arena for exploring various conjectures.
(I)

\item \label{IA} ``Physics with large extra dimensions: String Theory under experimental test,'' I. Antoniadis, Current Science {\bf 81} (12), 1609-1613 (2001).  Antoniadis discusses recent ideas through which the extra dimensions (beyond the familiar four of length, width, depth, and time) required by string theory might, with some luck, soon become detectable through experiment.
(I)

\item \label{SH} ``Future Science,'' S. Hawking, Current Science {\bf 81} (12), 1614-1615 (2001).  While not about string theory {\it per se},
Hawking's speculations on the future are included here as they are part of {\it Current Science}'s special issue on String Theory.
(I)

\end{enumerate}

\subsection{Books}

\label{ugbook}

\begin{enumerate}[{\bf 1{.}}]

\addtocounter{enumi}{31}

\item \label{BZ} {\bf A First Course in String Theory}, B.
Zwiebach (Cambridge University Press, Cambridge, 2003).  
Zweibach makes an explicit attempt to be
accessible to undergraduate students.  No familiarity with
quantum field theory is assumed (though Zwiebach does assume a working knowledge of quantum {\it Mechanics}).   It is intended for the
advanced undergraduate or beginning graduate level, roughly comparable
to Goldstein's mechanics. Zwiebach focuses on the study of single
strings and their interactions, which can be understood in some
detail. Of necessity, however, it is impossible to address many
advanced topics.  Note that this book may not be available until some time in 2004. (I)

\end{enumerate}

\section{General}

\label{gen}

Here I first provide a broad overview of the important string-theory resources and then describe in more detail
resources that provide a general introduction to the subject.  This literature is at   
a higher level than that of section \ref{ele} and is ``advanced" by the standards of this Resource Letter, as it requires at least basic graduate-level courses in
quantum field theory and, in some cases, general relativity.   Reviews of more advanced topics are described in
sections \ref{first}-\ref{last}.

\subsection{Journals}

The major vehicle of communication for the string community is not a journal at all.  Instead, results are primarily exchanged through the hep-th archive maintained by arxiv.org [http://www.arxiv.org/].
I provide arxiv reference numbers (in the form hep-th/0311044, with hep-th perhaps replaced by hep-ph, gr-qc, or math.ag) for most articles listed below.  Papers are typically first posted to the arxiv  as ``preprints," and only
later submitted to a journal.  

However, in the end, papers are published in the journals listed below.
The order in which I listed them reflects my own best guess (based only on anecdotal data)
as to the fraction of string papers published in each journal.  However, owing to reliance on the above archive, journal
selection is less critical in this field than in others.  The distinctions implied below are not overly large and a given
author will typically have a favorite journal dictated largely by personal taste.

The journals I list as Primary Journals are the most common avenues of
publication.  All manner of results are published there, and
these journals would be of most importance for readers of this
Resource Letter.  However, readers should be aware that some
results of a more specialized nature and targeted at those at the
interface with mathematics or in the gravitational-physics
community will instead be published in one of the Mathematical or
Gravitational Journals listed below.  Some reviews are also
published in {\it Physics Reports}. Occasional articles are
published in other journals as well.

\medskip

\noindent Primary Journals

{\it Journal of High Energy Physics}

{\it Physical Review D}

{\it Physical Review Letters}

{\it Advances in Theoretical and Mathematical Physics}

{\it Journal of Cosmology and Astroparticle Physics}

{\it Nuclear Physics B}

{\it Physics Letters B}

\noindent Mathematical Journals

{\it Communications in Mathematical Physics}

{\it Journal of Geometry and Physics}

{\it Journal of Mathematical Physics}

\noindent Gravitational Journals

{\it Classical and Quantum Gravity}

\noindent Review Journals

{\it Physics Reports}

\subsection{Conference Proceedings}

The primary string-theory conference is the Strings series, each labeled by the year (e.g., Strings '03 occurred in 2003).
The interested student can track the overall development of the field by leafing through the proceedings.  However, since the articles
are short summaries targeted at researchers already actively working in the field, students will wish to begin with
one of the textbooks or review articles I describe below.

Among the most useful resources for students learning string theory are the proceedings of various summer schools.  These schools are typically intended for graduate students and attempt to convey advanced topics in a reasonable pedagogical manner.  The proceedings are designed to allow students not present at the school to reap similar benefits.  The Theoretical Advanced Study Institutes (TASI) at the University of Colorado, Boulder, are an extremely useful series of annual schools with a first-rate proceedings.   Other excellent series of schools include those sponsored by the International Center for Theoretical Physics (ICTP) in Trieste, Italy,
and by the European RTN Network on
"The quantum structure of spacetime and
the geometric nature of fundamental interactions."  Many other excellent schools occur on a less regular basis. Rather than address each school itself in this letter, I will describe the proceedings articles resulting from specific lectures under review articles.  The individual articles are usually available from arxiv.org.

\subsection{Books}

\label{text}

I list here the main modern string textbooks are listed and provide a short description of them.  My comments are only a first guide to the literature, and the prospective student will also benefit from reading on-line book reviews from other students.  Such reviews can be found, for example, on the websites of various internet booksellers.  Note that ref. (\ref{CFT}) does not address string theory {\it per se}, but provides a marvelous reference on the related subject of conformal field theory.

\begin{enumerate}[{\bf 1{.}}]

\addtocounter{enumi}{32}

\item \label{Joe1} {\bf String Theory, Vol. 1: An introduction to
the Bosonic String}, J. Polchinski (Cambridge University Press,
Cambridge, 1998). Polchinski's two-volume {\it String Theory} is
the most comprehensive text addressing the discoveries of the
superstring revolutions of the early to mid 1990s, which mark the
beginnings of ``modern" string theory.  It is currently the most popular text for graduate-level courses in string theory.  Volume 1 quantizes the
bosonic string and uses this setting to introduce T-duality and
D-branes without the complications of Fermions. (A)

\item \label{Joe2} {\bf String Theory, Vol. 2: Superstring Theory
and beyond }, J. Polchinski (Cambridge University Press, Cambridge,
1998). Polchinski uses the first three chapters of volume 2 to
introduce superstrings, but quickly moves ``beyond." The rest of
this book provides an introduction to nearly all ``modern" string
topics pre-dating the AdS/CFT duality.  In particular, it includes D-branes,
Orbifolds, Black Hole Entropy, and Mirror symmetry. (A)

\item \label{cvj} {\bf D-branes}, C. Johnson (Cambridge
University Press, Cambridge, 2003).  This recent publication is clearly
the most up-to-date string text available.  It is an
excellent resource, beginning with a pedagogical quantization of a
single string and moving on to address many modern topics such as
fractional branes, the enhancon mechanism, and dielectric branes.
Of course, a certain brevity is required to fit both introductory
and advanced material into a single-volume book, so that beginning
students may wish to use {\it D-branes} in conjunction with one of
the older texts. (A)

\item{\bf Introduction to Superstring Theory}, E. Kiritsis (Leuven University Press, Leuven, 1998).  Also available from the arxiv as
hep-th/9709062.  Kiritsis provides a pedagogical introduction starting with the point particle and the bosonic string, and 
then proceeds through conformal
field theory and the superstring.  Advanced topics treated include T- duality, anomalies, compactification and supersymmetry breaking, 
loop corrections, and non-perturbative dualities.  The student will find this text very useful for cementing an understanding of core concepts before pursuing other advanced material. (A)

\item {\bf Superstring Theory: Vol. 1: Introduction},
 M.B. Green, J.H. Schwarz, and E. Witten (Cambridge University Press, Cambridge, 1987).
Volume 1 is the first of a classic two-volume string text by founders in the field.    Though there are now many more modern
texts, {\it Superstring Theory} is still extremely useful to the beginning student,  containing many details that more recent texts
must skim through to discuss modern topics.  Volume 1 is concerned mainly with the free string, though it treats both bosonic and supersymmetric versions. (A)

\item {\bf Superstring Theory: Vol. 2: Loop Amplitudes, Anomalies and Phenomenology},
 M.B. Green, J.H. Schwarz, and E. Witten (Cambridge University Press, Cambridge, 1987).   Volume 2 addresses the details of loop
calculations, but also discusses low-energy effective theories (in particular, supergravity) and compactifications (especially on Calabi-Yau manifolds).   Because more recent texts provide only a somewhat less detailed treatment of such subjects, {\it Superstring Theory: Volume 2} remains a useful reference. (A)

\item {\bf Lectures on String Theory}, D. L\"ust and S. Theisen (Springer-Verlag, Berlin, 1989).
This very useful text features a short route to the superstring and includes an excellent treatment of the heterotic string.   (A)

\item {\bf Supersymmetric Gauge Field Theory and Strings}, 
D. Balin and A. Love (Institute of Physics Publishing, Philadelphia, 1994). The real theme of this book is supersymmetry, beginning with the basics of supersymmetry algebras, progressing through supersymmetric gauge theory and supergravity, and finally introducing the perturbative superstring in the last half of the book.
By limiting the scope in this way, they are able to address these subjects in more detail than some other texts. (A)

\item{\bf Quantum Field Theory of Point Particles and Strings}, B. Hatfield (Perseus Publishing, London, 1998).  An updated version of Hatfield's 1992 book, the first half quickly reviews quantum field theory while the second introduces the very basics of strings.  The result is
close to, but still somewhat above the ``intermediate" level of this Resource Letter.  (A)

\item{\bf Gauge fields and Strings}, A. M. Polyakov (Harwood Academic Publishers, New York, 1987).  This classic work uses a detailed treatment of quantum field theory in terms of first-quantized particles to make a natural transition to the first quantized string.  Details of path-integral measures and other fundamentals are well-presented. (A)

\item{\bf Introduction to Superstrings and M-Theory (Graduate Texts in Contemporary Physics)}, M. Kaku
(Springer Verlag, New York, 1999).  This updated version of an earlier text begins with the basics and, after some development, addresses advanced topics such as compactifications and M-theory.  As a one-volume text, it cannot be comprehensive and, as a result, has a rather light discussion of D-branes.  (A)

\item{\bf Strings, Conformal Fields, and M-Theory (Graduate Texts in Contemporary Physics)}, M. Kaku
(Springer Verlag, New York, 2000).   The intent of this text is to build on {\it Introduction to Superstrings and M-theory}, addressing a number of more modern topics and illustrating the relationship to a broad range of related topics.   (A)

\item \label{CFT} {\bf Conformal Field Theory}, P. Di Francesco, P. Mathieu, and D. Senechal (Springer Verlag, New York, 1997).  Conformal field theory is a critical technical tool in the study of strings and this encyclopedic reference is a favorite among string theorists. (A)

\item {\bf Introduction to Global Supersymmetry}, P. Argyres, unpublished lecture notes available at
http://www.physics.uc.edu/~argyres/661/index.html.  Argyres gives an excellent modern introduction to supersymmetry. (A)

\item {\bf Quantum Fields and Strings: A
Course for Mathematicians}, edited by P. Deligne, P. Etingof, D. S. Freed. L. C. Jeffrey, D. Kazhdan, J. W. Morgan, D. R. Morrison, and E. Witten (American Mathematical Society, Providence, 1999).
Available online at http://www.math.ias.edu/QFT/fall/ and http://www.math.ias.edu/QFT/spring/index.html.  These edited lectures were given as a set of courses on quantum field theory and string theory for practicing mathematicians at the Princeton Institute for Advanced Study. (A)

\end{enumerate}

\subsection{Review Articles}

I gather here and in the remaining sections a list of useful review articles, though of course it is far from complete. The
SPIRES High Energy Physics Literature Database maintained by the
Stanford Linear Accelerator Center (SLAC) is a wonderful resource
for locating additional reviews. SPIRES maintains a database of
current publications and arxiv submissions in high-energy physics.
Their listings are extremely comprehensive from the beginning of
the arxiv in 1991.  Many useful entries from
earlier decades can also be found. Their guide to review literature in high-energy
physics (http://www.slac.stanford.edu/spires/hep/reviews.shtml) is
compiled both from user suggestions and by automated searching
through their database for review articles with more than 50
citations.  However, as noted in their own guide, proceedings of
summer schools often receive few citations even though they are
extremely useful. Instructions for using SPIRES automated
searching to find additional reviews are also given on the
web site above.

Most review articles introducing string theory as a whole have now
either been eclipsed by or transformed into the textbooks I
describe above.  As a result, I list only one general review article, ref. (\ref{open}), here
and separate the rest by topic in the following sections. I have also included
Strassler's gauge-theory review, ref. (\ref{MS}), and Ginsparg's conformal field theory review, ref. (\ref{PG}), which, while they do not address string theory {\it per
se}, they describe important 
tools used in some of the more advanced material below.  The reader may also enjoy refs. (\ref{JS}), (\ref{TY}), and (\ref{RG}) in section \ref{ug}.  Finally, ref. (\ref{AV}) contains another useful list of references on string theory and related topics.

\begin{enumerate}[{\bf 1{.}}]

\addtocounter{enumi}{47}

\item \label{open} ``Open Strings,'' C. Angelantonj and A. Sagnotti, Phys.\ Rept.\  {\bf 371}, 1-150 (2002)
[Erratum-{\it ibid}.\  {\bf 376}, 339-405 (2003)]
[arXiv:hep-th/0204089]. While not the most accessible review, Angelantonj and Sagnotti cover a broad range of material concerning open strings.  In particular, this work is known as an important reference for the details of many orbifold and orientifold constructions.
(A)

 \item \label{MS} ``An unorthodox
introduction to supersymmetric gauge theory,'' M.~J.~Strassler, 
arXiv:hep-th/0309149.  While not addressing string theory {\it per se}, I include this recent review owing to its timeliness and because an understanding of supersymmetric gauge theory is an important prerequisite for many topics in string physics. (A)

\item \label{PG} ``Applied Conformal Field Theory,''
 P.~Ginsparg,
arXiv:hep-th/9108028, in {\bf Les Houches Summer School in Theoretical Physics, Session 49: Fields, Strings, Critical Phenomena, Les Houches, France, June 28 - August 5, 1988}, 
edited by E. Brezin and J. Zinn-Justin (North-Holland, Amsterdam, 1990).  
These lecture notes remain a valuable reference on conformal field theory and Kac-Moody algebras. (A)

\item \label{AV} ``Recommended Literature
on Fields, Strings, Symmetries
and Integrable Systems,'' A. Vladimirov, http://thsun1.jinr.ru/~alvladim/lit.html.  Vladimirov presents a useful list of references on strings and related topics. (A)
\end{enumerate}

\section{Branes and Black Holes}

\label{first}

The most exciting developments in string theory from the last decade
concern the physics of so-called ``branes." These higher-dimensional relatives of strings come in many forms, the most
famous of which are the D-branes that launched exploration into
non-perturbative aspects of stringy physics.  All branes are
closely related to black holes, and D-branes have been used to
provide a stringy accounting of black-hole entropy.  More
recently, many researchers have investigated the idea that our
universe itself may be a brane in a higher-dimensional space (though I have placed this sub-topic with string phenomenology in section \ref{braneworld}).

While branes are also introduced in
some textbooks in section
\ref{gen}, the reviews below provide additional information and, at times, a different perspective. 
In particular, students
familiar with general relativity may prefer to start with ref. (\ref{marolf}) or (\ref{stelle})
which introduce branes through analogies with black holes.

The reviews here primarily address the so-called ``BPS branes," which are stable in superstring theories.
Non-BPS branes (most of which are unstable) are also of significant interest but I have grouped these with their primary applications in section \ref{nonBPS}.  The reader may also benefit from the discussion of U-duality in ref. (\ref{Boris}) of section \ref{dual} and the less-technical discussion in ref. (\ref{SW}) insection \ref{ug} of how string theory reproduces properties of black holes.

\begin{enumerate}[{\bf 1{.}}]
\addtocounter{enumi}{51}

\item ``The Quantum Physics Of Black
Holes: Results From String Theory,''  S.~R.~Das and S.~D.~Mathur, Ann.\ Rev.\ Nucl.\ Part.\
Sci.\  {\bf 50}, 153-206. (2000) [arXiv:gr-qc/0105063]. This review takes the reader as directly as possible to the D-brane description of black-hole entropy and Hawking radiation.  (A)

\item ``Microscopic formulation of black holes in string theory,''
J.~R.~David, G.~Mandal and S.~R.~Wadia,
Phys.\ Rept,\  {\bf 369}, 549-686 (2002)
[arXiv:hep-th/0203048].  The authors review the calculation of black-hole properties from string theory for the case
of the nearly-extreme D1-D5 black hole.  Many important details of the D-brane gauge and conformal field theories are discussed, leading to black-hole thermodynamics and Hawking radiation.  (A)

\item ``TASI lectures on black
 holes in string
theory,'' A.~W.~Peet, arXiv:hep-th/0008241, in {\bf Strings,
Branes, and Gravity: TASI 99: Boulder, Colorado, 31 May - 25 June 1999}, edited by J. Harvey, S. Kachru, and E. Silverstein (World Scientific, Singapore, 2001), pp.  353-433. Peet's thorough introduction to black holes and branes takes the reader up though a D-brane treatment of black-hole entropy and Hawking radiation.  (A)

\item \label{marolf} ``String/M-branes for
relativists,'' D.~Marolf,  arXiv:gr-qc/9908045. My own brief introduction to
branes, intended to convey some basic aspects of brane physics and
perspectives on string theory to those trained in general
relativity.  (A)

\item \label{stelle} ``BPS branes in supergravity,'' K.~S.~Stelle, 
arXiv:hep-th/9803116, in {\bf 1997 summer school in high energy physics and cosmology}, edited by E. Gava {\it et al.} (World Scientific, New Jersey, 1998), pp. 29-127.
Stelle provides a useful introduction to brane solutions in supergravity, containing many details of
supersymmetry, Kaluza-Klein reduction, and low-velocity scattering of branes.  (A)

\item ``Black holes and solitons in string theory,''
 D.~Youm,
Phys.\ Rept,\  {\bf 316}, 1 (1999)
[arXiv:hep-th/9710046].
An encylclopedic review of black-hole and black-brane solutions.

\item ``Four lectures on M-theory,'' P.~K.~Townsend, 
arXiv:hep-th/9612121, in {\bf 1996 Summer School in High Energy Physics and Cosmology. ICTP, Trieste, Italy, 10 June-26 July 1996}, edited by E. Gava {\it et al.}, (World Scientific, New Jersey, 1997), pp. 385-438.
These lectures describe the unification of superstring theories by M-theory, concentrating on aspects of
superalgebras and properties of branes. (A)

\item ``Black holes in string theory,'' J.~M.~Maldacena, 
arXiv:hep-th/9607235.   Maldacena's Ph.D. thesis was not written as an introduction for outsiders, but does contain detailed treatments of gauge-theory aspects relevant to D-brane counting of black-hole entropy that are hard to find in other sources.  Other useful aspects of brane and black-hole physics are also described. (A)

\end{enumerate}

\section{Gauge/Gravity duality}
\label{dual}

The idea of ``duality" is that a given quantum system could have
two (or more) radically different descriptions.  In interesting
cases, each description takes a simple form in some region of
parameter space but these regions do not overlap.  Thus, in one
regime it is efficient to use one description, while in
another regime it is efficient to describe the system in a
completely different way.

In the early 1990s, the so-called S- and T-dualities allowed
string theorists to investigate non-perturbative connections
between the various string theories.  These dualities are well described in modern string textbooks, especially refs.
(\ref{Joe1}), (\ref{Joe2}), and (\ref{cvj}).  However, more recently 
even more surprising dualities were discovered.  
Although there are no complete proofs of the duality conjectures, I
use the term ``discovered'' owing to the weight of circumstantial evidence in their favor.
 
These new ``gauge/gravity
dualities" concerned systems that could be described both as
theories of quantum gravity {\it and} as {\it non}-gravitating
quantum gauge theories on a {\it fixed} spacetime background.
Interestingly, the gravitating and non-gravitating descriptions
involve different numbers of spacetime dimensions.  The most
well-known and well-studied examples are Matrix theory and the
AdS/CFT correspondence.  The latter has recently been of particular interest in the so-called ``planewave limit"
in which much more detailed calculations can be performed.  

I have listed some notable reviews of these subjects below.  
I have
also included ref. (\ref{link}),  which gives a useful description of how the classic work on $c=1$ matrix models
and two-dimensional string theory reviewed in refs. (\ref{c1}) and (\ref{c2}) fit into the framework of gauge/gravity duality.
Another special case is
ref. (\ref{dS}), which reviews speculations as to how such
dualities might be extended to de Sitter space.
The reader may also enjoy refs. (\ref{EW}) and (\ref{ST}) in section \ref{ug}.

\begin{enumerate}[{\bf 1{.}}]

\addtocounter{enumi}{59}

 \item  ``TASI lectures on matrix
theory,'' T.~Banks, arXiv:hep-th/9911068,
in {\bf Strings,
Branes, and Gravity: TASI 99: Boulder, Colorado, 31 May - 25 June 1999}, edited by J. Harvey, S. Kachru, and E. Silverstein (World Scientific, Singapore, 2001), pp.  495-542.  This review by a founder of matrix theory emphasizes discrete light-cone quantization and
compactifications.  (A)

\item ``M(atrix) theory: Matrix quantum mechanics as a
fundamental theory,'' W.~Taylor,  Rev.\ Mod.\ Phys.\  {\bf 73}, 419-462 (2001)
[arXiv:hep-th/0101126]. Taylor makes an effort to make his review accessible, to cover a broad range of topics, and to describe how matrix theory relates to other models.  (A)

\item \label{Boris} ``U-duality and M-theory,''
N.~A.~Obers and B.~Pioline,
Phys.\ Rept.\  {\bf 318}, 113-225 (1999)
[arXiv:hep-th/9809039].  While primarily a review of U-duality, this work includes an important discussion of dualities in matrix theory.
(A)

\item ``Large N field theories, string theory and gravity,''  O.~Aharony, S.~S.~Gubser, J.~M.~Maldacena, H.~Ooguri and
Y.~Oz, Phys.\ Rept.\  {\bf 323}, 183-386 (2000) [arXiv:hep-th/9905111]. This impressive 261-page work is the canonical review of ADS/CFT. It is the most comprehensive such review I know, and future reviews of more advanced topics will no doubt assume this material as a starting point.  Nevertheless, the student should not expect to find all arguments spelled out in detail; for this, reference is made to the original works.  The gauge-theory perspective is emphasized, but many supergravity issues are also well discussed.
(A)

\item``TASI lectures: Introduction to the AdS/CFT
correspondence,''  I.~R.~Klebanov, arXiv:hep-th/0009139, in {\bf Strings,
Branes, and Gravity: TASI 99: Boulder, Colorado, 31 May - 25 June 1999}, edited by J. Harvey, S. Kachru, and E. Silverstein (World Scientific, Singapore, 2001), pp.  615-650.
Much shorter and less comprehensive than Aharony, {\it et. al.}, Klebanov provides a useful first introduction to the AdS/CFT correspondence.  (A)

\item  ``Lectures on branes, black holes and anti-de
Sitter space,'' M.~J.~Duff, arXiv:hep-th/9912164, in {\bf Strings,
Branes, and Gravity: TASI 99: Boulder, Colorado, 31 May - 25 June 1999}, edited by J. Harvey, S. Kachru, and E. Silverstein (World Scientific, Singapore, 2001), pp. 3-125.  Duff reviews the supergravity aspects of AdS/CFT.  (A)

\item ``Lecture notes on holographic
renormalization,'' K.~Skenderis,  Class.\ Quant.\ Grav.\  {\bf 19}, 5849-5876 (2002)
[arXiv:hep-th/0209067].  Some details of the AdS/CFT dictionary known as ``holographic renormalization" are reviewed in detail.
(A)

\item
``TASI 2003 lectures on AdS/CFT,'' J.~M.~Maldacena,
arXiv:hep-th/0309246.  Another name for the AdS/CFT correspondence is the ``Maldacena conjecture."  Here Maldacena's recent TASI lectures introduce the AdS/CFT correspondence and in particular the plane-wave limit. (A)

\item ``The plane-wave / super Yang-Mills duality,''
D.~Sadri and M.~M.~Sheikh-Jabbari,
arXiv:hep-th/0310119.  This is the most recent review available of the plane-wave limit of AdS/CFT.
(A)

\item \label{planewave} ``Light-Cone String Field
Theory in a Plane Wave Background,'' M.~Spradlin and A.~Volovich, arXiv:hep-th/0310033.  Spradlin and Volovich review the plane-wave limit of AdS/CFT and describe how string field theory calculations of string interactions in 9+1 dimensional plane-wave spacetimes match interactions calculated using N=4 super Yang-Mills gauge theory in a 3+1 conformally flat spacetime.  The paper includes
an elementary introduction to light-cone string field theory. (A)

\item \label{plane2} J.~C.~Plefka,
``Lectures on the plane-wave string / gauge theory duality,''
arXiv:hep-th/0307101.  Plefka's recent review of the plane-wave gauge/gravity duality concentrates on the gauge-theory side of
the correspondence and introduces an effective formalism that reduces the gauge-theory computations to quantum mechanics.
Plefka again describes how string field theory calculations match on to gauge theory calculations.  An elementary introduction to light-cone string field theory is included here as well. (A)

\item 
``Strings in plane wave backgrounds,'' A.~Pankiewicz,
arXiv:hep-th/0307027.  This fairly detailed introduction to the plane-wave limit of AdS/CFT is based on Pankiewicz's Ph.D. thesis. (A)

\item ``Spinning strings and AdS/CFT duality,'' A. Tseytlin, arXiv: hep-th/031113, to appear in Ian Kogan Memorial Volume, {\bf From   Fields to Strings: Circumnavigating Theoretical Physics}, edited by M. Shifman, A.   Vainshtein, and J. Wheater  (World Scientific, Singapore, 2004).  Tseytlin reviews how certain string states in Anti-de Sitter space (crossed with $S^5$) can be studied in detail using semiclassical methods, allowing detailed studies of the AdS/CFT duality to be carried out.  These states have large angular momentum on the $S^5$ and their study is related to the above plane-wave limit. (A)

 \item \label{c1}
``String theory in two-dimensions,'' I.~R.~Klebanov,
arXiv:hep-th/9108019.  This is a classic review of $c=1$ matrix models of two-dimensional string theory.  The modern perspective on this subject is described in ref. (\ref{link}) below, but the student may benefit from the more thorough introduction given here. (A)

\item \label{c2}  ``Lectures On 2-D Gravity And 2-D String Theory,'' P.~Ginsparg and G.~W.~Moore,
arXiv:hep-th/9304011.  Another classic review of $c=1$ matrix models and two-dimensional string theory.  While somewhat
more recent than ref. (\ref{c1}) above, it still pre-dates the modern perspective described in
ref. (\ref{link}) below.
(A)

\item  \label{link} ``D-brane decay in two-dimensional string theory,''
I.~R.~Klebanov, J.~Maldacena and N.~Seiberg,
JHEP {\bf 0307}, 045 (2003)
[arXiv:hep-th/0305159].  This paper is not a review, but provides a useful description of the modern perspective on $c=1$ matrix
models and their connection to D0-branes and gauge/gravity duality.
(A)

\item  \label{dS} ``Les Houches lectures on de Sitter space,'' M.~Spradlin, A.~Strominger, and A.~Volovich, arXiv:hep-th/0110007, in
{\bf Les Houches 2001, Gravity, gauge theories and strings}, 
edited by C. Bachas, A. Bilal, M. Douglas, N. Nekrasov and F. David
(Springer-Verlag, New York, 2002), pp. 423-453.
This work reviews speculations as to how gauge/gravity
dualities might extend to de Sitter space.  Useful background on de Sitter space is included.
(A)

\end{enumerate}

\section{String Field Theory, non-commutative geometry, Non-BPS branes, and tachyon condensation}
\label{nonBPS}

String field theory has seen a recent resurgence owing to its use in describing the decay of unstable branes.  Some researchers expect this to lead to a generalization of the gauge/gravity dualities above that would allow all of closed string theory to be described using only open strings.  The field continues to be an active form of research, but has now stabilized enough for some good reviews to be available.  Because tachyon condensation is the main current application of both string field theory and non-BPS branes, reviews of such subjects sometimes overlap strongly.  Another entangled subject is non-commutative geometry, which is again useful in the study of tachyon condensation.  As a result, I have grouped all of these subjects together in this section.  The reader interested in particular aspects of this work will receive useful guidance from the annotations below.  See also ref. (\ref{AS}) for a less technical discussion,  refs. (\ref{planewave}) and (\ref{plane2}) in section \ref{dual}, which also introduce string field theory, but focuss on its applications to the plane-wave limit of AdS/CFT, and ref. (\ref{Q}) in section \ref{braneworld}, which describes possible cosmological
applications of decaying and rolling tachyons.

\begin{enumerate}[{\bf 1{.}}]
\addtocounter{enumi}{76}

\item ``D-Branes, Tachyons, and String Field Theory,''
W.~Taylor and B.~Zwiebach,
arXiv:hep-th/0311017. This long-awaited review will no doubt become the standard modern introduction to string field theory.
Taylor and Zwiebach consider the open bosonic string and build their review around the study of tachyon condensation and what non-perturbative information can be obtained from string field theory.  
(A)

\item 
``Topics in string field theory,'' L.~Bonora, C.~Maccaferri, D.~Mamone, and M.~Salizzoni, arXiv:hep-th/0304270. This more-technical review of open bosonic string field theory again describes the construction of D-branes and the associated tachyon decay. (A)

\item ``A review on tachyon condensation in open string field theories,'' K.~Ohmori,
arXiv:hep-th/0102085.  Based on Ohmori's master's thesis, this is a very comprehensive review of open string field theory and tachyon decay as of early 2001. Both the bosonic and supersymmetric cases are discussed. (A)

\item {\bf Introduction To String Field Theory}, W.~Siegel (World Scientific, Singapore, January 1989).  Also,
arXiv:hep-th/0107094.  Siegel's  textbook on string field theory provides the most complete review available of classic pre-1989
material.  Of course, since it dates from before the recent string field theory renaissance, it does not address many modern topics.
Siegel recently posted the book to the arxiv from which the student can download the complete text free of charge.  (A)

\item 
``Non-BPS states and branes in string theory,'' A.~Sen,
arXiv:hep-th/9904207, in {\bf Progress in string theory and M-theory: proceedings of the NATO Advanced Study Institute on Progress in String Theory and M-Theory, Cargese, France, May 24-June 5, 1999}, edited by Laurent Baulieu, {\it et al}, (Kluwer Academic, Amsterdam, 2001), pp. 187-234.
Sen's conjectures for unstable branes were the motivating force behind the recent research in tachyon decay.  Here Sen reviews the basics of non-BPS branes and their associated tachyons, including how one brane can be described as a tachyon soliton of another brane. (A)

\item ``TASI lectures on non-BPS D-brane systems,'' J.~H.~Schwarz, 
arXiv:hep-th/9908144,
in {\bf Strings,
Branes, and Gravity: TASI 99: Boulder, Colorado, 31 May - 25 June 1999}, edited by J. Harvey, S. Kachru, and E. Silverstein (World Scientific, Singapore, 2001), pp.  809-846.  Schwarz reviews a wide variety of both stable and unstable non-BPS branes. (A)

\item 
``Lectures on non-BPS Dirichlet branes,'' M.~R.~Gaberdiel,
Class.\ Quant.\ Grav.\  {\bf 17}, 3483-3520 (2000)
[arXiv:hep-th/0005029].  Gaberdiel provides a complete introduction to the boundary-state formalism and its applications to the construction of both BPS and non-BPS branes. (A)

\item \label{VL1}
``D branes in string theory. I,'' P. Di Vecchia and A. Liccardo,
in {\bf M-theory and quantum geometry},  edited by L. Thorlacius and T. Jonsson
(Kluwer Academic Publishers,  Boston, 2000).
This review also describes the boundary-state construction in detail.  (A)

\item
``D-branes in string theory. II,'' P.~Di Vecchia and A.~Liccardo,
arXiv:hep-th/9912275.  Building on ref. (\ref{VL1}), more complicated properties of boundary states are reviewed.  The Dirac-Born-Infeld action is derived, and the boundary-state construction of a particular stable non-BPS D-brane is described.
(A)

\item ``Stable non-BPS states in string theory: A pedagogical review,'' A. Lerda and R. Russo,
Int. J. Mod. Phys. {\bf A 15}, 771-820 (2000)
[arXiv:hep-th/9905006].  Lerda and Rousso provide a pedagogical review of stable non-BPS branes and emphasize their relations under various dualities.  Tachyon condensation is discussed both qualitatively and in detail through conformal field theory. (A)

\item ``Noncommutative field
theory,'' M.~R.~Douglas and N.~A.~Nekrasov,  Rev.\ Mod.\ Phys,\  {\bf 73}, 977-1029 (2001)
[arXiv:hep-th/0106048]. Field theory on a non-commutative spacetime is introduced and reviewed.  Care is paid to mathematical details, but topics are introduced with an eye to their use in string theory.  Applications to quantum Hall systems, Matrix theory, and AdS/CFT are discussed.  (A)

\item ``Komaba lectures on noncommutative solitons
and D-branes,'' J.~A.~Harvey,  arXiv:hep-th/0102076.  Harvey introduces the basics of non-commutative geometry and non-commutative field theory, focussing on the construction of solitons and applications to D-branes. (A)

\end{enumerate}

\section{String Phenomenology, Brane Worlds, and Orbifolds}
\label{braneworld}

String phenomenology is the attempt to link string theory to the
real world by reproducing the standard model of particle physics, cosmology,
or other significant aspects of our universe.  Perhaps the most profound mysteries in this context concern
the cosmological constant, which appears to be small but non-zero in our universe.  Much of the recent
excitement has been in the so-called ``brane-world" context, in
which our universe is a 3+1 dimensional version of a
membrane embedded in a higher-dimensional space.  The Calabi-Yau manifold references in section (\ref{math}) are also useful in the study of phenomenology and, in addition, 
the student may enjoy the undergraduate-level description of brane-worlds in ref. (\ref{IA}) in section \ref{ug}.
I have grouped these subjects together with stringy orbifold compactifications because the latter are important tools in the study of 
string models (both with and without brane worlds).  Note that ref. (\ref{open}) and several textbooks in section \ref{gen} 
are also useful orbifold resources.

The reader should be aware of four more specialized topics also described in the works below.  First, note that there has been much recent discussion as to how to construct stringy models that lead to 
inflation (i.e., de Sitter-like spacetimes) in four dimensions.  As this subject is still too immature for there to be
useful reviews, I have instead included refs. (\ref{KKLT}) and (\ref{KKLT2}) which represent the state of the art in this field.
A second special topic is the discussion in
(\ref{Ruan}) that addresses more mathematical aspects of orbifolds.  The third concerns the so-called orientifolds, which are
a sub-category of orbifolds that can have negative tension and, in particular, they are the only
gravitating negative-energy objects generally believed to be self-consistent.  Because they are somewhat more technical, however, 
they are rarely treated in detail in reviews (see ref. (\ref{open}) in section \ref{gen} and ref. (\ref{EK}) below for exceptions) or textbooks.  I have included refs. (\ref{orient1}) and (\ref{orient2}) below to help the interested reader bridge this gap.  See also ref. (\ref{orient3}).
Finally, ref. (\ref{time}) reviews the modern subject of ``time-dependent orbifolds," which has little to do with traditional phenomenology but is a useful arena for exploring fundamental issues associated with spacelike singularities. This review also ties in to various cosmological
issues discussed by other works in this section.

As described here, string phenomenology is an extremely broad subject without clear boundaries to allow sharp divisions into sub-fields.  To impose some element of order, I have attempted to place reviews which concentrate on particle physics toward the beginning of the list, followed by those concentrating on cosmology, and finally those describing
more technical aspects of orbifolds.  
However, any such separation is imperfect and imprecise at best.  The student may find more useful guidance from the annotations
below.

\begin{enumerate}[{\bf 1{.}}]
\addtocounter{enumi}{88}

\item \label{EK} ``D-branes in standard model building, gravity and cosmology,''
 E. Kiritsis, hep-th/0310001.
This recent and extremely broad review describes many aspects of D-branes and related objects such as orbifolds and orientifolds with emphasis on phenomenological applications.  Supersymmetry breaking, axions, model building, warped extra dimensions, and cosmological implications are discussed.
(A)

\item ``TASI lectures on M theory phenomenology,'' M.~Dine, 
arXiv:hep-th/0003175, in {\bf Strings,
Branes, and Gravity: TASI 99: Boulder, Colorado, 31 May - 25 June 1999}, edited by J. Harvey, S. Kachru, and E. Silverstein (World Scientific, Singapore, 2001),
pp. 545-612. Dine provides important perspective and
 background for the student of string phenomenology.  He reviews
 both the issues to be addressed and
the tools to address them.  Issues discussed include the cosmological
constant, the hierarchy problem, CP violation, the strong CP
problem, and many others.  Tools described include supersymmetry,
supersymmetry breaking, string moduli, and others.  The outsider
will find this a very useful introduction. (A)

\item ``Heterotic orbifolds,'' J.~T.~Giedt,  arXiv:hep-ph/0204315.
Based on Giedt's Ph.D. thesis, this work addresses the
reproduction of the standard model through techniques not
involving brane worlds.  Giedt begins with a useful review of
orbifold constructions and of the heterotic string, and then
proceeds to illustrate how a certain class of orbifold
compactifications can lead to standard-model-like effective
theories in 4 dimensions. (A)

\item ``Orbifold Compactifications Of
String Theory,'' D.~Bailin and A.~Love, Phys.\ Rept.\  {\bf 315}, 285-408 (1999).  Bailin
and Love provide a thorough review of orbifold compactifications
from the perspective of a string phenomenologist.  They begin with
the basics of orbifolds, and then move on to address model-building issues such as Yukawa couplings, K\"ahler potentials, and
supersymmetry breaking. (A)

\item K.~R.~Dienes,
``Neutrinos from strings: A Practical introduction to string theory, string model building, and string phenomenology. I: Ten-dimensions,'' 
{\bf Neutrinos in physics and astrophysics : from ${\bf 10^{-33} cm}$ to ${\bf 10^{+28} cm}$ ; TASI 98, Boulder, Colorado, USA, 1-26 June 1998}, edited by Paul Langacker, (World Scientific, Singapore, 2000), pp. 201-302.   This is one of the few reviews of string model-building
from the phenomenological perspective.  The interested reader should watch for an expanded and updated version to be posted on the arxiv in 2004.
(A)

\item ``The second string (phenomenology) revolution,''
L.E. Ibanez, hep-ph/9911499,  Class. Quant. Grav.  {\bf 17}, 1117-1128 (2000).
It was once thought that the Heterotic string was the only reasonable candidate string theory for 
model-building.  Ibanez briefly reviews how this changed in the mid 1990s as a better understanding of the theory led to many new opportunities.
(A)

\item ``A review of three-family grand unified string models,'' Z. Kakushadze, G. Shiu, and S.-H.H. Tye,
        hep-th/9710149, Int. J. Mod. Phys. {\bf A13}, 2551-2598 (1998).
This review provides a thorough description of the so-called ``asymmetric-orbifold" models that lead to grand unification and
three families of particles. (A)

\item ``String Theory and the Path to Unification: A Review of Recent Developments,''
K.R. Dienes, hep-th/9602045, Phys. Reports {\bf 287}, 447-525 (1997).
This thorough pedagogical review describes the state of string phenomenology as of 1996, with an eye
toward unification and supersymmetry considerations.
(A)

\item ``Lectures on superstring phenomenology,'' F. Quevedo, hep-th/9603074, in {\bf Workshops on particles and fields and phenomenology of fundamental interactions, Puebla, M\'exico, November 1995}, edited by J. C. D'Olivo, A. Fern\'andez, and M. A. P\'erez 
(AIP Press, Woodbury, N.Y., 1996), pp. 202-242.
String phenomenology is described as of 1996, concentrating on effective Lagrangians, supersymmetry breaking, and dualities. (A)

\item ``Large dimensions and string physics in future colliders,'' I. Antoniadis and K. Benakli, hep-ph/0007226, Int. J. Mod. Phys. 
{\bf A15}, 4237-4286 (2000). This review describes a number of large-extra-dimension scenarios with low string scales.
Open issues and observational bounds are discussed. (A)

\item \label{Q} ``Lectures on string / brane cosmology,'' F. Quevedo, hep-th/0210292, Class. Quant. Grav. {\bf 19}, 5721-5779 (2002). 
This review describes the status of brane-world cosmology as of 2002.  Particular consideration is given to inflation or alternatives. Modern topics such as negative-tension branes, S-branes, and rolling tachyons are discussed.  (A)

\item ``ICTP lectures on large extra dimensions,'' G. Gabadadze, hep-ph/0308112.
This pedagogical review begins with Kaluza-Klein theory and then introduces the various flavors of brane-worlds, including both large-extra-dimension and warped-compactification scenarios.   The emphasis is on cosmology, with particular interest in explaining the observed acceleration of our universe. (A)

\item ``Brane worlds,'' R.~Dick,  Class.\ Quant.\ Grav.\  {\bf 18},
R1-R24 (2001) [arXiv:hep-th/0105320]. Dick provides an introduction to
brane worlds focusing on gravitational issues and cosmology.
Emphasis is on the construction of gravitating brane-world
solutions.  (A)

\item  ``Cosmology and brane worlds: A
review,'' P.~Brax and C.~van de Bruck, Class.\ Quant.\ Grav.\  {\bf 20}, R201-R232 (2003)
[arXiv:hep-th/0303095].  This review considers rane worlds and
their applications to cosmology.  Multiple brane scenarios are
included and brane collisions are briefly discussed. (A)

\item \label{KKLT}
``De Sitter vacua in string theory,'' S.~Kachru, R.~Kallosh, A.~Linde and S.~P.~Trivedi,
Phys.\ Rev.\ D {\bf 68}, 046005 (2003)
[arXiv:hep-th/0301240].  While not a review, this paper, together with ref. (\ref{KKLT2}), represents the current state of the art with respect to 
finding string compactifications that yield inflating four-dimensional spacetimes. (A)

\item \label{KKLT2}
``Towards inflation in string theory,'' 
S.~Kachru, R.~Kallosh, A.~Linde, J.~Maldacena, L.~McAllister and S.~P.~Trivedi,
JCAP {\bf 0310}, 013 (2003)
[arXiv:hep-th/0308055]. See comments for ref. (\ref{KKLT}).

\item \label{Ruan}  ``Stringy orbifolds,'' Y.~b.~Ruan, arXiv:math.ag/0201123.
Ruan discusses the mathematical aspects of orbifolds, including
cohomology and twisted K-theory. (A)

\item \label{orient1} ``On orientifolds, discrete torsion, branes and M theory,''
A.~Hanany and B.~Kol,
JHEP, {\bf 0006}, 013 (2000)
[arXiv:hep-th/0003025].  While not a review, this paper provides a useful discussion of orientifold planes,  
focussing on M-theory descriptions.
(A)

\item \label{orient2} ``Orientifolds, RR torsion, and K-theory,''
O.~Bergman, E.~G.~Gimon and S.~Sugimoto,
JHEP {\bf 0105}, 047 (2001)
[arXiv:hep-th/0103183].  While also not a review paper, the first part contains a good review of orientifolds.
The rest discusses the role of Ramond-Ramond fluxes from the K-theory point of view. (A)

\item \label{orient3}  ``Triples, fluxes, and strings,''
J.~de Boer, R.~Dijkgraaf, K.~Hori, A.~Keurentjes, J.~Morgan, D.~R.~Morrison and S.~Sethi,
Adv.\ Theor.\ Math.\ Phys.\  {\bf 4}, 995 (2002)
[arXiv:hep-th/0103170].  This sizeable review addresses string compactifcations with

\item \label{time}  ``Time-dependent orbifolds and string cosmology,''
L.~Cornalba and M.~S.~Costa,
arXiv:hep-th/0310099.  Time-dependent orbifolds have been an active area of research over the past few years.
Interesting examples are simple models in which one can explore certain consequences of spacelike singularities.
In a certain sense, they also approximate the big-bang singularity as described in the ekpyrotic and cyclic models of cosmology.

\end{enumerate}

\section{Mirror Symmetry and Calabi-Yau manifolds}
\label{math}

Calabi-Yau manifolds are a particularly well-studied type of string compactification.  Such compactifications are extremely useful in string phenomenology, but also form a sub-field in their own right.  This sub-field often has a strongly mathematical flavor and is a favorite of researchers near the boundary between physics and mathematics.  Mirror Symmetry, a form of duality that arises in Calabi-Yau manifolds, has been a topic of particular interest.  

\begin{enumerate}[{\bf 1{.}}]
\addtocounter{enumi}{109}
\item {\bf Mirror Symmetry}, edited by K. Hori, S. Katz, A. Klemm, R. Pandharipande, R. Thomas, C. Vafa, R. Vakil, and E. Zaslow (Americal Mathematical Society, Providence, 2003).    Hori {\it et al} present the proceedings of the Clay Mathematics Institute School on Mirror Symmetry held at Harvard in June, 2000.  The collected lectures provide an excellent introduction to two-dimensional N=2 supersymmetric field theory, mirror symmetry, D-branes on Calabi-Yau manifolds, and other mathematical aspects of string theory.  (A)

\item ``String theory on Calabi-Yau manifolds,''
B.~R.~Greene,
arXiv:hep-th/9702155 in
{\bf Fields, strings, and duality : TASI 96 : Boulder, Colorado, US, 2-28 June 1996} edited by  C. Efthimiou and  B.~R.~Greene (World Scientific, Singapore, 1997), pp. 543-726.
Greene provides a thorough introduction to Calabi-Yau manifolds and stringy geometry.

\item \label{MSI} {\bf Mirror Symmetry I}, edited by S. T. Yau (American Mathematical Society, Providence, 1998).  Along with ref. (\ref{MSII}), this text is an updated addition of the first collection of papers published after the discovery of mirror symmetry.  Thus the papers in this volume describe their subjects without assuming prior knowledge of mirror symmetry, though they do assume familiarity with 
Calabi-Yau manifolds.  (A)

\item \label{MSII} {\bf Mirror Symmetry II}, edited by B. Greene and S. T. Yau (American Mathematical Society, Providence, 1997).  See comments for ref. (\ref{MSI}).
(A)

\item {\bf Mirror Symmetry and Algebraic Geometry}, D. Cox and S. Katz (American Mathematical Society, Providence, 1999).  This volume distills the mathematics, and in particular the algebraic geometry, of mirror symmetry from the literature.  toric varieties, Todge theory, KŠhler geometry, moduli of stable maps, Calabi-Yau manifolds, quantum cohomology, Gromov-Witten invariants, and such are described in a style aimed at mathematicians.
(A)

\item ``Compactification, geometry and duality: N = 2,'' P.~S.~Aspinwall, 
arXiv:hep-th/0001001, in {\bf Strings,
Branes, and Gravity: TASI 99: Boulder, Colorado, 31 May - 25 June 1999}, edited by J. Harvey, S. Kachru, and E. Silverstein (World Scientific, Singapore, 2001),
pp. 723-805.  Aspinwall reviews the geometry of the moduli space of string compactifications with N=2 supersymmetry, with attention to differences between such cases and those with more supersymmetry. 
(A)

\item ``K3 surfaces and string duality,''
P.~S.~Aspinwall,
arXiv:hep-th/9611137, in {\bf Fields, strings, and duality : TASI 96 : Boulder, Colorado, US, 2-28 June 1996} edited by  C. Efthimiou and  B.~R.~Greene (World Scientific, Singapore, 1997), pp. 543-726.
Greene provides a thorough introduction to Calabi-Yau manifolds and stringy geometry.
These lectures address heterotic compactifications on K3 and the dual formulations in the IIA and IIB theories, emphasizing geometrical aspects of K3. (A)

\end{enumerate}

\section{The holographic principle}
\label{last}

In this last section I address the so-called ``holographic principle,"
a set of ideas that some researchers believe to be fundamental to
string theory or any theory of quantum gravity.  However, this
claim remains controversial and the detailed relationship of these
ideas to string theory has not yet been established.  I include the subject
here because it forms a lively and fascinating area of
discussion and research.  As these discussions typically do not rely on
special string knowledge (and can in fact be followed at a rough level
with only minimal knowledge of quantum field theory and general
relativity), I have rated these as ``intermediate" on the scale
described in the introduction.  The reader may also enjoy ref. (\ref{ST}) in section \ref{ug}.

The student should be aware that both sides of this discussion are
not represented equally below.  I know of no proper review dedicated to arguing
for caution in the use of holography.  In place of such a review, I
have included refs. (\ref{Wald}) and (\ref{obs}) below.

\begin{enumerate}[{\bf 1{.}}]

\addtocounter{enumi}{116}

\item \label{BS} ``TASI lectures on
the holographic principle,'' D.~Bigatti and L.~Susskind,  arXiv:hep-th/0002044,
in {\bf Strings,
Branes, and Gravity: TASI 99: Boulder, Colorado, 31 May - 25 June 1999}, edited by J. Harvey, S. Kachru, and E. Silverstein (World Scientific, Singapore, 2001), pp. 883-933.
This short
review describes the basic ideas behind the holographic principle
and tests of the idea in AdS/CFT.   The work also discusses some
aspects of black-hole information loss and Susskind's proposal of
``black-hole complementarity." (I)

\item \label{RB} ``The holographic principle,'' R.~Bousso,  Rev.\
Mod.\ Phys.\ {\bf 74}, 825-874 (2002) [arXiv:hep-th/0203101]. Bousso
provides the most thorough review I know of holography and related
issues. A number of applications and examples are presented and
some relevant black-hole physics is reviewed. (I) \newline

\item \label{Wald} ``The thermodynamics of black holes,'' R. Wald, 
Living Rev.\ Rel.\  {\bf 4}, 6 (2001) [arXiv:gr-qc/9912119]. Wald
provides a thorough review of black-hole thermodynamics, a subject
fundamental to many discussions of holography.  While not a review
of holography {\it per se}, some interesting comments on holographic and
related ideas can be found near the end of this work. (I)

\item \label{obs} ``On the status of
highly entropic objects,'' D.~Marolf and R.~Sorkin,  arXiv:hep-th/0309218.  While not a
review, this rather accessible recent work is included to provide
further contrast with refs. (\ref{BS}) and (\ref{RB}) above. (I)

\end{enumerate}

\acknowledgements
I thank the participants of the 2003 KITP Superstring Cosmology program and especially Marcus Berg, Keith Dienes, Jerome Gauntlett, Alberto Lerda, Per Kraus, JR Minkel, Shiraz Minwalla, Boris Pioline, Jan Plefka, Joe Polchinski, Radu Roiban, Savdeep Sethi, Ted Sung, Alexei Vladimirov, Spenta Wadia, Johannes Walcher, and Peter Woit for useful comments and suggestions.  I gratefully acknowledge partial support from NSF grants PHY99-07949
and PHY03-54978 and the University of California.

\end{document}